\documentstyle[12pt]{article}

\def\bi{\bigskip}
\def\be{\begin{equation}}
\def\en{\end{equation}}
\def\bq{\begin{eqnarray}}
\def\eq{\end{eqnarray}}
\def\noi{\noindent}
\begin{document}

\begin{center}
{\Large \bf HEAVY BARYON SPIN 3/2 THEORY AND RADIATIVE DECAYS OF THE DECUPLET\footnote{Supported by CONACyT under contract 4918-E}}\\[1.5cm]
\end{center}
\vspace{1cm}
\begin{center}
{{\large \bf M. Napsuciale and J.L. Lucio M.}\\[1cm]
\vspace{.8cm}
{\it Instituto de F\'\i sica, Universidad de Guanajuato}\\
{\it Apartado Postal E-143, Le\'on, Gto., M\'exico}}
\end{center}
\noindent

\vspace{2cm}

\begin{center}
{\bf Abstract.} 
\end{center}

We study the radiative decays of the decuplet $\bigg(\frac{3}{2}\to \frac{1}{2}\gamma\bigg)$ using Heavy Baryon Chiral Perturbation Theory (HBChPT). We
emphasize the problems faced by the interacting spin $\frac{3}{2}$
field theory. We argue that, to lowest ~order ~in ~the $\frac{1}{m}$ 
~expansion, ~HBChPT ~provides ~a ~framework ~where ~R-invariance and the 
appropiated constraints for the interacting spin $\frac{3}{2}$ fields are 
consistently incorporated. We perform a gauge invariant calculation of the 
decay amplitudes, to lowest order in $\bigg( \frac{1}{m} \bigg)$ and to order 
$\frac{w^2}{\Lambda^2_\chi}$ in the chiral expansion and report analytical 
results for the one-loop contributions to the two form factors involved in the
$\frac{3}{2} \to \frac{1}{2} \gamma$ transitions. Parameters independent 
predictions for the SU(3) forbidden decays are presented.

\setlength{\baselineskip}{1\baselineskip}

\newpage

Chiral Perturbation Theory is a useful tool with wich to describe ~low 
~momentum processes involving Goldstone Bosons [1]. When the formalism is 
extended to interactions with Baryons (BCHPT) it suffers from two unpleasant 
characteristics: i) there is no correspondence between the loop and momentum 
expansion and ii) the expansion parameter turn out to be of order one. A new 
formalism (HBCHPT) was ~proposed ~by ~Jenkins ~and ~Manohar (J-M) wich 
circunvents these problems [2]. The new theory is written in terms of definite
velocity baryon fields for which the Dirac equation co\-rres\-ponds to a 
massless baryon. The Heavy Baryon Lagrangian is expressed in terms of a 
$\frac{1}{m}$ expansion. The $\frac{1}{m}$ effects in HBCHPT are considered in
HBCHPT by the inclusion of higher dimension operators suppressed by inverse 
powers of m. 

\bi

In addition to the baryon octet, the baryon decuplet is also included [3]. In 
this case the theory involves spin 3/2 fields whose effective mass is $\Delta 
m =M-m$ (m-M masses of the baryon octet and decuplet respectively). It is 
possible to construct an effective theory by integrating out the decuplet 
fields. Virtual effects of the decuplet in the theory are then ~considered by 
higher dimension ~operators ~involving ~only ~octet ~baryons ~and ~mesons 
which are suppressed by inverse powers of $\Delta m$. In the real world the 
decuplet-octet baryon mass difference $\Delta m \approx 300 MeV$ is small when
compared with the hadronic scale of 1GeV. The smallness of $\Delta m$ produces
an enhancement of the decuplet effects, therefore it is advantageous to retain 
explicity decuplet fields in the effective theory, rather than integrate them 
out. 

\bi

The formalism previously described has been recently used by Butler, Savage 
and Springer (BSS) [4] to carry out a detailed study of the radiative 
decays of the decuplet $(T\to B \gamma)$. The scheme is promising as it 
provides a systematic field theoretical approach which is able to produce 
testable predictions. From our point of view, the analysis of BSS can be 
improved if the following points are addressed:

\bi

\noi - ~It is well known that Quantum Field Theory for spin 3/2 interacting 
fields suffers of serious inconsistencies. In particular, within the 
Rarita-Schwinger formalism [6] the R-S spinor $\psi_\mu$ has more degrees of 
freedom than required, which results in a non-unique classical Lagrangian. In 
fact there exist a whole family of one-parameter Lagrangians ${\cal L}(A)$ 
from which the Dirac-Fierz-Pauli equation and the necessary free field 
constraints for $\psi_\mu$ may be obtained [7]. Furthermore, these 
Lagrangians remain invariant under point transformations (R-invariance) which 
mixes the spurious spin 1/2 fields contained in $\psi_\mu$. This invariance 
~ensures ~the ~unphysical ~spin ~1/2 ~degrees ~of ~freedom ~have ~no 
~observable ~effects. ~Generalizations ~of ~this ~scheme ~to ~describe 
~interacting spin 
3/2 ~fields requires that the co\-rres\-pon\-ding Lagrangian preserves 
R-invariance. ~This introduces new ambiguities into the theory (the so-called 
``off-shell" ~parameters) ~[8,12]. ~It ~should ~also ~be ~emphasized ~that the 
~constraint ~equations, ~which ~achieve ~the ~elimination ~of ~the ~unphysical
spin 1/2 degrees of freedom in the free case, are no longer valid for 
interacting fields. As shown below, only the leading order of the 
$\frac{1}{m}$ expansion is free of the ``off-shell" parameters ambiguities.

\bi

\noi - ~The $\frac{1}{m}$ and the chiral $\bigg( \frac{w}{\Lambda_\chi}\bigg)$ 
expansions must be separately considered. An interacting spin 3/2 theory which 
is free of the ``off-shell" ambiguities limitates to the zeroth order the
$\bigg(\frac{1}{m}\bigg)$ expansion. On the chiral the expansion must be 
consistently
carried, order by order, including all possible counterterms.

\bi

\noi - ~Existing calculations for $T\to B\gamma$ amplitudes have been 
performed in the $\epsilon \cdot v = 0$ gauge. We understand that a non-gauge 
invariant calculation is valid whenever it is carried in a consistent way. 
However, a gauge invariant analysis is desirable as it allows an unambiguous 
determination of the counterterms required, the form factors and thereby of 
the multipole amplitudes $E1$ and $M2$.

\bi

The paper ~is organized ~as ~follows. ~In ~section 1 we ~summarize ~the 
~kinematics and the invariant amplitudes including the counterterms and the 
heavy baryon limit. In section 2 we incorporate the electromagnetic 
interactions to the approach used in [12] to describe the R-invariant 
interacting spin 3/2 fields. In section 3 we consider Heavy Baryon Chiral 
Perturbation Theory (HBChPT). The R-invariant Lagrangians as well as the 
subsidiary conditions on the spin $3/2$ field are considered in the light of 
the $\frac{1}{m}$ expansion. Section 4 is devoted to the radiative decay of 
the decuplet, details of the calculations are presented as our analytical 
results differ from previous calculations [4]. Further details on Chiral 
Perturbation Theory and $SU(3)$ Clebsch-Gordon coefficients are deferred to an
appendix.

\newpage

\noi {\bf 1.- GENERAL CONSIDERATIONS}

\bi
\bi

Lorentz covariance, gauge invariance and parity dictates the more general form 
of the invariant amplitude describing the $3/2^+ \to 1/2^+ \gamma$ 
transitions. For on-shell external baryons the amplitude is parametrized as:

\be
i {\cal M}_{fi} = e \bar\psi (q) \Gamma_{\mu\nu} \psi^\mu (p) \varepsilon^\nu ,
\en

\noi where the vertex tensor $\Gamma_{\mu\nu}$ can be written as:

\be
\Gamma_{\mu\nu}= \frac{g^{^D}_{_1}}{2\Lambda} (k_\mu \gamma_\nu - \not k 
g_{\mu\nu}) \gamma_5+ \frac{g^{^D}_{_2}}{4\Lambda^2} ~\frac{1}{M+m} (q 
\cdot k g_{\mu\nu} - k_\mu q_\nu) \not k \gamma_5 ,
\en

\noi $\Lambda$ is a still unspecified mass scale and $M(m)$ is the $3/2^+ 
(1/2^+)$ baryon mass. This is essentially the characterization used  
in [9], the only differences are that we use the mass scale
$\Lambda$ instead of the spin $1/2$ baryon mass $m$ to normalize the form 
factors, and $p$ is written in terms of $q$ by using the equations of motion.  

\bi

\noi The processes we are considering gets contributions from magnetic dipole 
$(M1)$ and electric quadrupole $(E2)$ amplitudes. In terms of the
form factors $g^{^{D}}_{_{1}} , g^{^{D}}_{_{2}}$ the multipole amplitudes are 
given by:

\bq
M1 &=& \frac{e}{12\Lambda} \bigg(\frac{w}{Mm} \bigg)^{1/2} \bigg( 
(3M+m) g^{^D}_{_1} - \frac{M(M-m)}{2\Lambda} g^{^D}_{_2} \bigg) \nonumber \\
E2 &=& - \frac{e}{6\Lambda} ~\frac{w}{M+m} \bigg( \frac{wM}{m}\bigg)^{1/2} (g^{^D}_{_1} - \frac{M}{2\Lambda} g^{^D}_{_2} \bigg)
\eq

\noi where $w$ is the photon energy in the rest frame of the $3/2^+$
particle. The decay width is expressed in terms of the multipoles:

\be
\Gamma (T\to B \gamma)= \frac{w^2}{2\pi} ~\frac{m}{M} \left\{ | M1|^2 + 3 |E2|^2 \right\} .
\en

\noi Our purpose in this paper is to work out the predictions of HBChPT,
which requires considering the heavy baryon limit of (1) and (2). This is 
achieved by considering the baryons as fields of definite velocity.
Following [3] we write $M=m +\Delta m; ~q=mv+\ell$ and take the $m \to \infty$
limit to obtain 

\be
\Gamma^{HB}_{\mu\nu} = P_+ \Gamma_{\mu\nu} P_+ = \frac{g^{^D}_{_1}}{2\Lambda}
(k_\mu S_\nu - S\cdot k g_{\mu\nu})+ \frac{g^{^D}_{_2}}{8\Lambda^2} (v\cdot k
g_{\mu\nu} -k_\mu v_\nu) S\cdot k
\en

\noi where $S_\mu$ is the heavy baryon limit of the spin operator

\be
S_\mu = \frac{i}{2} \gamma_5 \sigma_{\mu\rho} v^\rho ,
\en

\noi and $P_+$ stand for the proyector over the positive energy subspace (see 
Eq. (20) below).

\bi

Radiative decays of the decuplet  are induced by 1-loop diagrams which are 
generated with the leading order (dimension 4) HBChPT Lagrangian. Actual 
calculation (see (24,26) below) shows that the 1-loop induced amplitude is 
${\cal O} \bigg( \frac{w^2}{\Lambda^2_\chi} \bigg)$, hence besides the leading 
order chiral Lagrangian, three counterterms are required since to ${\cal O} 
\bigg( \frac{w^2}{\Lambda^2_\chi}\bigg)$ one dimension five and two dimension 
six counterterms contribute to the decuplet radiative decays at tree level.

\bq
{\cal L}^1_{_{CT}} &=& e \frac{\Theta_1}{\Lambda_\chi} \bar B S^\theta Q T^\mu F_{\theta\mu} \nonumber \\
{\cal L}^2_{_{CT}} &=& -ie \frac{\Theta_2}{\Lambda^2_\chi} \bar B v^\alpha S^\theta Q T^\mu \partial_\alpha F_{\theta\mu} \\
{\cal L}^3_{_{CT}} &=& ie \frac{\Theta_3}{\Lambda^2_\chi} \bar B v^\theta S^\alpha Q T^\mu \partial_\alpha F_{\theta\mu} \nonumber
\eq

\noi These counterterms reproduce the Lorentz structure indicated in (5). 
$\Theta_1$ is finite whereas $\Theta_2$ and $\Theta_3$ renormalize the 
one-loop contributions. 

\bi
\bi

\noi {\bf 2.- CHIRAL SPIN 3/2 THEORY}

\bi

We will not review ~the ~formalism ~for ~Chiral ~Perturbation ~Theory 
~involving only spin 1/2 and pseudoscalar fields (for a review see [3]). 
Instead we  will focus our attention on the spin 3/2 sector as we believe the 
longly known problems faced by spin 3/2 field theory have not been properly 
discussed in the light of chiral symmetry and the Heavy Field approximation.  

\bi
\bi

\noi {\bf a) Free fields.}

\bi

\noi Within the Rarita-Schwinger approach, the S=3/2 field is described by a 
spinor-vector $\psi_\mu$ which is obtained from the tensor product of a spinor
and a four-vector. Clearly $\psi_\mu$ involves more degrees of freedom than
required. This redundancy is reflected in the formalism in the need for the 
subsidiary conditions:

\begin{eqnarray}
\gamma^\mu \psi_\mu (x) &=& 0 \nonumber \\
\partial^\mu \psi_\mu (x) &=& 0 .
\end{eqnarray}

\bi

\noi It has been shown that for the spin 3/2 field, there exist a whole family
of one parameter Lagrangians from wich the equation of motion and the
subsidiary conditions (1) can be derived [7]

\bi

\begin{equation}
{\cal L} (A) = \psi^\mu (x) \{i \partial_\alpha \Gamma^\alpha\,_{\mu\nu} (A) -
M B_{\mu\nu} (A) \} \psi^\nu (x)
\end{equation}

\noi where

\begin{eqnarray}
\Gamma^\alpha_{\mu\nu}(A) &=& g_{\mu\nu} \gamma^\alpha + B \gamma_\mu 
\gamma^\alpha \gamma_\nu + A (\gamma_\mu g^\alpha\,_\nu + g_\mu\,^\alpha
\gamma_\nu ) . \nonumber \\
B_{\mu\nu} (A) &=& g_{\mu\nu} - C \gamma_\mu \gamma_\nu \nonumber \\
A \not = -1/2 &,& B=\frac{3}{2} A^2 +A+\frac{1}{2} , ~~C= 3A^2+3A+1 . \nonumber
\end{eqnarray}

\bi

\noi For A=-1/3, ${\cal L} (A)$ reduces to the Lagrangian originally proposed 
by R-S [6].

\bi
\bi

\noi By construction the Lagrangian (9) is invariant under the point 
transformations (R-invariance):

\bi

\begin{equation}
\psi_\mu \to \psi^\prime_\mu = R_{\mu\alpha} (a) \psi^\alpha \qquad\qquad
A \to A^\prime = \frac{A-2a}{1+4a}
\end{equation}

\bi

\noi where

\bi

$$R_{\mu\nu} (a) = g_{\mu\nu} + a \gamma_\mu \gamma_\nu \qquad\qquad 
a \not = - \frac{1}{4}$$

\bi

The R operator acts only on the spin 1/2 components of $\psi_\mu$. The 
arbitrariness of the spin 1/2 components of $\psi_\mu$ is at the origin
of the family of one parameter Lagrangians (9).

\bi

Notice the factorization property (which in fact can be generalized to the
lagrangian including interaction).

\bq
\Gamma_{\mu\nu}^\alpha (A) &=& R_\mu\,^\beta (h(A)) \Gamma^\alpha_{\beta\rho} 
(A=-\frac{1}{3}) R^\rho\,_\nu (h(A)) \nonumber \\
B_{\mu\nu} (A) &=& R_\mu\,^\beta (h(A)) B_{\beta\rho} (A=- \frac{1}{3} ) R^\rho\,_\nu (h (A)) \nonumber
\eq

\noi with $h(A) = \frac{1}{2} (1+3A)$. Useful properties of the R operator are

\begin{eqnarray}
R_{\mu\nu} (a) R^\nu\,_\lambda (b) &=& R_{\mu\lambda} (a+b+4ab); \nonumber \\
R^{-1}_{\mu\nu} (a) &=& R_{\mu\nu} \bigg(- \frac{a}{4a+1} \bigg) , \\
R_{\mu\nu} (0) &=& g_{\mu\nu} . \nonumber
\end{eqnarray}

\bi
\bi

\noi The generalization of (9) to include interactions should lead to physical
quantities which are A-independent. On the other hand, the Kamefuchi-O'Raifeartaigh-Salam (K.O.S.) theorem [10] assures the A-independence of the S matrix
elements if the whole Lagrangian ${\cal L} = {\cal L}_{free} + {\cal L}_{int}$
is R-invariant. Therefore once we work with an R-invariant Lagrangian we are 
free to use any value of A, physical quantities being A-independent.

\bi
\bi

\noi {\bf b) Interacting Fields.}

\bi

Chiral symmetry has been traditionally taken as the guiding principle to 
construct phenomenological Lagrangians aiming to describe the existing data 
[11]. Further requirements for any sensible Lagrangians involving spin 3/2 
fields are:

\bi

\noi - ~~~R-Invariance. The full Lagrangian, ${\cal L} = {\cal L}_{free} +
{\cal L}_{int}$ must be invariant under contact R-transformations. This
ensures, through the K-O-S theorem [10] the A independence of physical
quantities.

\bi

\noi - ~~~Constraints. Once the interactions are included, the Lagrangian 
must lead to the appropiated subsidiary conditions so that only the correct 
number of degrees of freedom to describe a spin 3/2 field are left.

\bi

A convenient procedure to construct R-invariant Lagrangian describing 
interacting spin 3/2 fields was discussed in [12]. To lowest order in the 
chiral expansion, the Lagrangian is

\begin{equation}
{\cal L} = {\cal L}_{free} + {\cal L}_{\frac{3}{2}\frac{1}{2}0} + {\cal L}_{\frac{3}{2}\frac{3}{2}0} . 
\end{equation}

\bi

The subindices $\frac{3}{2} \frac{1}{2} 0$ and $\frac{3}{2}\frac{3}{2}0$ stand
for the spin of the fields included in the corresponding Lagrangian. The free 
Lagrangian is given in (9) and the R-invariant interactions read

\begin{eqnarray}
{\cal L}_{\frac{3}{2}\frac{1}{2}0} &=& i {\cal C} \bar\psi^\mu O_{\mu\nu} 
(A,Z)\psi \Delta^\nu + h.c. \nonumber \\
{\cal L}_{\frac{3}{2}\frac{3}{2}0}&=&i{\cal H} \bar\psi^\mu O_{\mu\nu\alpha}
(A,X,Y) \psi^\nu \Delta^\alpha 
\end{eqnarray}

\bi

\noi ${\cal C}$ and ${\cal H}$ are coupling constants and $\Delta_\mu$ is the
axial-vector chiral covariant field given in (A4) of the appendix. The vertex 
tensors can be written as [8,12]

\bi

\bq
{\cal O}_{\mu\nu} (A,Z) &=& (g_{\mu\nu}+f(A,Z) \gamma_\mu \gamma_\nu) \\
{\cal O}_{\mu\nu\alpha} (A,X,Y) &=& R_{\mu\beta} (f (A,X) g^\beta\,_\sigma 
\gamma_\alpha \gamma_5 R^\sigma\,_\nu (f(A, Y)) 
\eq

\noi with

\be
f(A,V)= \frac{1}{2} (1+4V) A+V \qquad\qquad\quad V= X,Y,Z 
\en

\noi where $X,Y,Z$, are arbitrary (``off-shell") parameters. Here and 
thereafter we omit the SU(3) structure. Details on these vertices as well as 
the Clebsch-Gordon coefficients are given in the appendix. 

\bi

We still need to derive the $T-B-\gamma$ R-invariant interaction, which has the
following Lorentz structure 

\be
{\cal L}_{\frac{3}{2}\frac{1}{2}\gamma} = e \bar \psi^\mu{\cal O}^{(\gamma)}_{\mu\alpha\beta} (A) \gamma_5 \psi F^{\alpha\beta} ,
\en

\bi
\bi

\noi ${\cal O}^{(\gamma)}_{\mu\alpha\beta} (A)$ is splitted as

$${\cal O}^{(\gamma)}_{\mu\alpha\beta} =\frac{1}{2} g_1 {\cal O}^{(1)}_{\mu\alpha\beta}+ \frac{1}{2} g_2 {\cal O}^{(2)}_{\mu\alpha\beta}$$

\noi with

\bq
{\cal O}^{(1)}_{\mu\alpha\beta}&=& R_{\mu\lambda} (\ell (A))(g^\lambda\,_\alpha
\gamma_\beta - g^\lambda\,_\beta \gamma_\alpha) \nonumber \\ 
{\cal O}^{(2)}_{\mu\alpha\beta} &=& R_{\mu\lambda} (\ell (A)) (g^\lambda\,_\alpha \partial_\beta - g^\lambda\,_\beta \partial_\alpha )  \nonumber
\eq

\noi $\ell (A)$ is, in principle, an arbitrary function of A, however the 
requirement of R-invariance for the Lagrangian restricts $\ell (A)$ to the 
class of functions for which the following identify holds

$$\ell (A) = (1+4a) \ell (A^\prime) + a$$

\noi For linear functions $\ell (A)$, the solution is given by (16) where now 
$V$ stand for $U,W$; two new arbitrary parameters. Therefore, the R  invariant 
$T-B-\gamma$ interaction reads:

\bq
{\cal L}_{\frac{3}{2}\frac{1}{2}\gamma} &=& e \bar\psi^\mu \{ g_1 
(g_{\mu\alpha} \gamma_\beta + f (A, W) \gamma_\mu \gamma_\alpha \gamma_\beta)
+ \nonumber \\ 
& & g_2 (g_{\mu\alpha} \partial_\beta + f (A,U) \gamma_\mu \gamma_\alpha 
\partial_\beta) \} \gamma_5 \psi F^{\alpha \beta} .
\eq

\bi

Summarizing, R-invariance wich ensures the A-independence of physical 
quantities, can be implemented in the Lagrangian at the price of introducing 
arbitrary ``off-shell" parameters. Sometimes these ``off-shell" parameters are
arbitrarily fixed, thus for example in [13] R-invariance is implemented by 
impossing the condition $\gamma^\mu O_{\mu\nu} =0$ which amounts to pick up
the value $Z=-1/4$. Other authors use the ``off-shell" parameters to fit the 
data  [9,11]. 

\bi

In adittion to R-invariance we need adequate subsidiary conditions
in order to eliminate the redundant degrees of freedom contained in 
$\psi_\mu$. In the following section we will discuss the subsidiary conditions 
in the light of the Heavy Baryon approach, wich is required in order to get a
sensible chiral expansion. 

\bi
\bi

\noi {\bf 3.- HEAVY BARYON SPIN 3/2 THEORY}

\bi

When the decuplet is introduced in the theory where the octet baryons are 
treated as heavy fields, the spin $\frac{3}{2}$ fields of definite velocity 
are defined by:

$$\psi^\mu_v (x) = e^{imv.x} \psi^\mu (x)$$

\noi where $m$ stand for the nucleon mass in the chiral limit, {\it i.e.} the 
decuplet Lagrangian contains an explicit decuplet mass term proportional to 
$\Delta m = M -m$. 

We assumme that the HBChPT is defined by the Feynman rules which are obtained
from the $\frac{1}{m}$ expansion of the lagrangian derived in the last
section.

In the heavy baryon limit $(m \to \infty)$, writing $P_\mu =m v_\mu +\ell_\mu$ 
and keeping leading terms in the $\frac{1}{m}$ expansion, the free lagrangian 
(9) leads to the propagator [12]: 
  
\begin{eqnarray}
i\Delta_{\mu\nu} (P,A)= \bigg\{ &-&g_{\mu\nu} + \frac{1}{3} \gamma_\mu \gamma_\nu - \frac{1}{3} (\gamma_\mu v_\nu - v_\mu \gamma_\nu )\qquad\qquad\qquad \nonumber \\
&+&\frac{2}{3} v_\mu v_\nu \bigg\} \frac{P_+}{(v \cdot k-\Delta m+i\epsilon)} 
+ {\cal O} \bigg( \frac{1}{m} ; A \bigg) \nonumber 
\end{eqnarray}
\be
\equiv i \Delta^0_{\mu\nu} +{\cal O} \bigg( \frac{1}{m} ; A \bigg)\qquad\qquad\quad\qquad\qquad\quad
\en

\noi where

\be
P_+  = \frac{1+ \not v}{2}
\en

\noi It is straigthforward to check that

\begin{equation}
\gamma^\mu \Delta^0_{\mu\nu} = \Delta^0_{\mu\nu} \gamma^\nu = 0
\end{equation}

The important point is ~that  (21) implies that, to zeroth order in the 
$\frac{1}{m}$ expansion, terms in the Lagragian containing the ``off-shell" 
parameters do not contribute to S-matrix elements. ~Indeed,  ~any 
~calculation regarding the Lagrangian (12), involves the vertices $O_{\mu\nu}$
and $O_{\mu\nu\alpha}$ wich connect either to an external on-shell heavy 
baryon or to a heavy baryon propagator $\Delta^0_{\mu\nu}$. The constraint for
the free field (1), or relation (20) for the heavy propagator ensure in both 
cases the vanishing of the contributions arising from ``off-shell" parameters 
in the Lagrangian.

\bi

These results suggests that to leading order in the $\frac{1}{m}$ expansion, 
the cons\-traints $\gamma^\mu\psi_\mu =0$ do not get modified by the 
interacting terms. In fact, in [12] it was shown that modifications to
the free 
constraints due to interactions are ${\cal O} \bigg(\frac{1}{m} \bigg)$, 
therefore, the interacting spin 3/2 field theory is free of ``off-shell" 
ambiguities only to order zero in the $\frac{1}{m}$ expansion. Furthermore, to
that order in the $\frac{1}{m}$ expansion the elimination of the redundant 
degrees of freedom is guaranteed since the free field constraints still hold. 

\bi

To order zero in the $\frac{1}{m}$ expansion we can just forget the 
``off-shell" terms, in whose case the interaction reduces to the one commonly 
used in the li\-te\-ra\-tu\-re [3]. It is important to remark however that we 
get a different free Lagrangian and even more important, our analysis shows 
that beyond the zeroth order in the $\frac{1}{m}$ expansion the 
``off-shell" ambiguities appear and mo\-di\-fi\-ca\-tions to the subsidiary 
conditions due to the interactions have to be reconsidered. Consequently
we need to keep track separately of the heavy baryon and chiral expansions, 
even if the corresponding parameters are of the same order of magnitude, m the
baryon mass $\approx$ 1 GeV, and $\Lambda_\chi$ $\approx$ 1 GeV.

\newpage

\noi {\bf 4.- $\frac{3}{2} \to \frac{1}{2} + \gamma$ DECAYS}

\bi

Previous calculations of the $T \to B \gamma$ decays in the framework of 
HBChPT were carried by BSS [4]. The present work is an improvement of such
work as we address the problems of the interacting spin $\frac{3}{2}$ field 
theory and we perform a  gauge invariant calculation wich permits to 
unambiguously identify the form factors. In particular we have argued in the 
previous section that, it we work to order zero in the $\frac{1}{m}$ 
expansion, HBChPT provides a framework to incorporate interacting spin 
$\frac{3}{2}$ fields which is both R-invariant and free of the so called 
``off-shell"  ambiguities. 

\bi

According to these results, below we calculate the amplitude for the $T \to
B \gamma$ decay to lowest order in $\bigg( \frac{1}{m} \bigg)$ and to order
$\frac{w^2}{\Lambda^2_\chi}$ in the chiral expansion. This is in contrast with
BSS calculation [4] where both expansion are mixed under the argument that
$m\approx\Lambda_\chi$. Futhermore, once we decide to work to ${\cal O} \bigg( 
\frac{w^2}{\Lambda^2_\chi} \bigg)$ in the chiral expansion, the Lorentz 
structure of the amplitude (5) indicates that we require 3 counterterms. One 
finite ${\cal O} \bigg(\frac{w}{\Lambda_\chi}\bigg)$ and two ${\cal O} 
\bigg(\frac{w^2}{\Lambda^2_\chi}\bigg)$, which arise naturally once a gauge 
invariant calculation is performed. 

\bi
\bi

\noi {\bf 4a) Loop Calculations}

\bi

For the processes under consideration there are twelve generic diagrams. Using 
the Feynman rules derived from the Heavy Baryon lagrangian (summarized in the 
appendix) it is possible to show that the only non vanishing contribution arise
from those diagrams shown in fig. (1). As our analytical results differs from 
those given in [4] and the differences turn out to be numerically relevant, 
below we give some details on the calculations. For diagram (a) we obtain

\be
{\cal M}_a =- \frac{2e}{f^2} C_{_{TBB}} \bar B S^\theta T^\mu 
\varepsilon^\nu \tau_{\theta\mu\nu}
\en

\noi where

$$\tau_{\theta\mu\nu} = \int \frac{d^4 \ell}{(2\pi)^4} ~\frac{(\ell + k)_\theta \ell_\mu \ell_\nu}{((\ell+p)\cdot v+i \xi) ((\ell +k)^2 - m^2_p+i\xi) (\ell^2 -m^2_p+i\xi)}$$

\bi

\noi $m_p$ denotes the mass of the pseudoescalar entering in the loop and 
$C_{TBB}$ stand for the products of Clebsch-Gordon coefficients involved in a 
particular process. Using Feynman's parametrization and dimensional 
regularization this expression reduces to

\bq
\tau_{\theta\mu\nu} &=&\frac{i}{(4\pi)^2}(4\pi \mu^2)^\varepsilon \int^1_0 ~dx
\int^\infty_0 ~d\lambda\bigg\{\Gamma(\varepsilon)\frac{(1-x)g_{\mu\nu}k_\theta
-xg_{\nu\theta} k_\mu}{(\lambda^2 +2\lambda\beta +\gamma)^\varepsilon} \\
&-& \Gamma (1+\varepsilon ) \frac{2\lambda x (1-x)}{(\lambda^2 +2\lambda \beta
+ \gamma)^{1+\varepsilon)}} k_\theta k_\mu v_\nu \bigg\} \nonumber
\eq  

\bi
\noi where $\varepsilon \equiv 2 - \frac{d}{2}, \mu$ is the mass scale 
introduced by dimensional regularization, $\beta \equiv (k\cdot v) ~x- (p\cdot
v)$ and ~$\gamma \equiv m^2_p -i\xi$.

\bi

\noi Using the identity

\bi

$$\frac{d}{dx} ~\frac{x(1-x)}{(\lambda^2 + 2\lambda \beta +\gamma)^\varepsilon} =\frac{1-2x}{(\lambda^2 + 2\lambda \beta+ \gamma)^\varepsilon} - \varepsilon \frac{2\lambda x (1-x) (k \cdot v)}{(\lambda^2 + 2\lambda \beta + \gamma)^{1+\varepsilon}}$$ 

\bi

Eq. (22) can be cast in an explicitly gauge invariant form

\bq
\tau_{\theta\mu\nu} &=& \frac{i}{(4\pi)^2} \int^1_0 dx \frac{1}{k \cdot v} 
\int^\infty_0 d\lambda \frac{(4\pi \mu^2)^\varepsilon \Gamma (\varepsilon)}{(\lambda^2 +2 \lambda \beta + \gamma)^2} \nonumber \\
&~& \bigg\{ (k\cdot v) \bigg( (1-x) g_{\mu\nu} k_\theta - x g_{\nu\theta} 
k_\mu \bigg) -(1-2x) k_\theta k_\mu v_\nu \bigg\} . \nonumber
\eq

\bi

\noi Eq. (22) can be rewritten in terms of two Lorentz and gauge invariant 
independent structures:

\be
i{\cal M}_a = \frac{e}{(4\pi f)^2} \bar B \{ g^a_1 I^{(1)}_{\mu\nu\theta} + 
g^a_2 I^{(2)}_{\mu\nu\theta} \} S^\theta T^\mu \varepsilon^\nu
\en

\bi
\noi where

\bq
I^{(1)}_{\mu\nu\theta} &=& k \cdot v ( g_{\nu\theta} k_\mu - g_{\mu\nu} 
k_\theta ) \nonumber \\
I^{(2)}_{\mu\nu\theta} &=& (k \cdot v  g_{\mu\nu} k_\theta - k_\theta 
k_\mu v_\nu )  
\eq

\bq
g^a_1 &\equiv & -2  (C G)_{_{TBB}} \int^1_0 ~xdx (4\pi \mu^2_1)^\varepsilon \Gamma (\varepsilon ) I(-\varepsilon , b, c) \nonumber \\
g^a_2 &\equiv & 2 (C G)_{_{TBB}} \int^1_0 ~(1-2x) dx  (4\pi \mu^2_1)^\varepsilon \Gamma (\varepsilon ) I(-\varepsilon , b, c) \nonumber 
\eq

\bi

\noi with

$$I(-\varepsilon ,b,c)\equiv \int^\infty_0 ~dx (\lambda^2 +2\lambda b+c)^{-\varepsilon} .$$

\bi

$$\mu_1 \equiv \frac{\mu}{k\cdot v} ~, \qquad b\equiv \frac{\beta}{k\cdot v}
=x- \frac{p\cdot v}{k\cdot v} , \qquad c\equiv \frac{\gamma}{(k\cdot v)^2} .$$

\bi

Before presenting the results of diagram (1b) the following comments are in 
order:

\begin{itemize}
\item[-] In the $v \cdot \varepsilon =0$ gauge the last term in the curly 
	 brackets of (23) is absent and the existence of two independent
	 gauge invariant structures is masked [4].
\bi
\item[-] The Feynman integration parameter in (23) is dimensionfull, which
	 obscures the power counting analysis. It proofs to be convenient to 
	 work with the dimensionless integration variable $\lambda /k\cdot v$.
 
\end{itemize}

\bi

\noi The second diagram is handled similarly, the only point to keep in mind is
that the spin $\frac{3}{2}$ has an effective mass $\Delta m= M-m$,

\be
i{\cal M}_b = \frac{e}{(4\pi f)^2} \bar B \{ g^b_1 I^{(1)}_{\mu\nu\theta} 
+ g^b_2 I^{(2)}_{\mu\nu\theta} \} S^\theta T^\mu \varepsilon^\nu
\en

\noi where

\bq
g^b_1 &=& -2 (CG)_{_{TTB}} \int^1_0 \bigg( 1 - \frac{x}{3} \bigg) ~dx (4\pi \mu^2_1)^\varepsilon \Gamma (\varepsilon )I (-\varepsilon ,b^\prime, c)\nonumber\\
g^b_2 &=& -2 (CG)_{_{TTB}} \int^1_0 ~\frac{1}{3} (1-2x) dx (4\pi \mu^2_1)^\varepsilon \Gamma (\varepsilon )I (-\varepsilon , b^\prime , c)\nonumber
\eq

\bi

\noi where $b^\prime =x-\frac{p\cdot v}{k\cdot v}+ \frac{\Delta m}{k\cdot v}$ 
and $(CG)_{_{TTB}}$ stand for the product of Clebsh-Gordon coefficients 
entering in diagram (b). Notice in this sense that the $b^\prime -b$ 
difference appearing in $g^\alpha_i; ~i=1,2; ~\alpha =a,b$ arises from the 
baryon propagators in the loop. Whereas fig. (1a) involves a massles baryon, 
in fig. (1b) a decuplet baryon of mass $\Delta m$ is propagating.

\bi

Using the recursion relations worked out by J-M [3] for the integrals 
$I(a,b,c)$ we obtain

$$(4\pi \mu^2_1)^\varepsilon I(-\varepsilon , b,c)=-b \bigg(\frac{1}{\varepsilon} -\gamma+ \ln 4\pi \bigg) +b \ln \bigg( \frac{c}{\mu^2_1} -2 \bigg) - 2 (c-b^2)I(-1,b,c) + {\cal O} (\varepsilon) .$$
 
\bi

\noi Within the modified minimal substraction scheme, the last equation is 
divided into a divergent and a finite part.

$$(4\pi \mu^2_1)^\varepsilon I(-\varepsilon ,b,c) = - b \bigg( \frac{1}{\varepsilon} -\gamma + \ln 4\pi\bigg) +I_{fin} (b,c)$$

\noi with

$$I_{fin} (b, c)=b \ln \bigg(\frac{c}{\mu^2_1}-2 \bigg) -2(c-b^2) I(-1, b,c)$$
    
\bi

In the rest frame of the spin 3/2 particle $p\cdot v=\Delta m$, $k\cdot v=w$. 
On the other hand $w=\Delta m + {\cal O} \bigg(\frac{1}{m} \bigg)$ so that to 
order zero in $\frac{1}{m}$, $w=\Delta m$. As the $\frac{1}{m}$ corrections 
for $w$ generates $\frac{1}{m}$ corrections for $g^\alpha_i$ we have $b^\prime
= x$; $b= x-1$.  

\bi

Combining the contributions of fig. (1) we get $(g^L_i = g^a_i + g^b_i)$

\bq
g^L_1 &=& \{ -2(CG)_{_{TBB}} \int^1_0 (1-x) I_{fin}(-x, c) \nonumber \\ 
&-& 2 (CG)_{_{TTB}} \int^1_0 (1-\frac{x}{3}) I_{fin} (x, c) \} \\
g^L_2 &=& \{ -2(CG)_{_{TBB}} \int^1_0 (1-2x) I_{fin} (-x, c)\nonumber \\
&-& 2 (CG)_{_{TTB}} \int^1_0 \frac{1}{3} (1-2x) I_{fin} (x, c) \} . \nonumber 
\eq

\bi

Using the t$^\prime$Hooft \& Veltman [14] conventions to deal with the 
logarithm branchs points of $I(-1,b,c)$ we calculate the finite part as:

\[ I_{fin} (x,c) = x \bigg( \ln \frac{m^2_p}{\mu^2} -2\bigg) + \left\{ \begin{array}{lll} \sqrt{x^2-\rho^2} ~\ln \bigg| \frac{\eta_+}{\eta_-} \bigg| & 
x > \rho \\
-2 \sqrt{\rho^2 - x^2} ~\mbox{arc ~tg} \left[ \frac{\rho^2 - x^2}{x^2} \right]^{1/2} & x < \rho \qquad\quad
\end{array}
\right. \]
\be
\en
\[ I_{fin} (-x, c) = -x \bigg( \ln \frac{m^2_p}{\mu^2} -2\bigg) + \left\{ \begin{array}{lll} -\sqrt{x^2-\rho^2} \bigg( ~\ln \bigg| \frac{\eta_+}{\eta_-} \bigg| - 2 \pi i \bigg) & x > \rho \\
2 \sqrt{\rho^2 - x^2} \bigg( ~\mbox{arctg} \sqrt{\frac{\rho^2}{x^2} -1} - \pi \bigg) & x < \rho \qquad
\end{array}
\right. \]

\bi

\noi where

\bi

\begin{equation}
\eta_\pm = x \pm \sqrt{|x^2 - \rho^2|} \qquad\qquad \rho \equiv\frac{m_p}{w} .
\end{equation}

\bi

Explicit calculations lead to  the analytical results reported in table 1.

\newpage

\begin{centerline}
{\bf Table 1}
\end{centerline}

$$\Delta \to N \gamma$$
$$g^{L}_{i} = 2 ~\frac{{\cal C}}{\sqrt{3}} \bigg\{ (F+D) \ell^a_i (\pi) - (F-D) \ell^a_i (K) - \frac{1}{3} {\cal H} (5 \ell^b_i (\pi) + \ell^b_i (K) )\bigg \}$$

$$\Sigma^{*0} \to \Lambda \gamma$$
$$g^{L}_{i} = 2 ~\frac{{\cal C}}{3} \bigg\{ -D (\ell^a_i (K) +2 \ell^a_i (\pi)) + {\cal H} (\ell^b_i (K) + 2 \ell^b_i (\pi) )\bigg \}$$

$$\Sigma^{*0} \to \Sigma^0 \gamma$$
$$g^{L}_{i} = 2 ~\frac{{\cal C}}{\sqrt{3}} \bigg\{ D \ell^a_i (K) - {\cal H}  ~\ell^b_i (K) )\bigg \}$$

$$\Sigma^{*+} \to \Sigma^+ \gamma$$
$$g^{L}_{i} = 2 ~\frac{{\cal C}}{\sqrt{3}} \bigg\{ - (D-F) \ell^a_i (\pi) - (D+F) \ell^a_i (K) + \frac{{\cal H}}{3} (\ell^b_i (\pi) +5\ell^b_i (K) )\bigg \}$$

$$\Sigma^{*-} \to \Sigma^- \gamma$$
$$g^{L}_{i} = 2 ~\frac{{\cal C}}{\sqrt{3}} \bigg\{ -(D-F) (\ell^a_i (\pi) - \ell^a_i (K)) + \frac{1}{3} {\cal H} (\ell^b_i (\pi) - \ell^b_i (K) )\bigg \}$$

$$\Xi^{*0} \to \Xi^0 \gamma$$
$$g^{L}_{i} = 2 ~\frac{{\cal C}}{\sqrt{3}} \bigg\{ -(D-F) \ell^a_i (\pi) - (D+F) \ell^a_i (K)+ \frac{1}{3} {\cal H} (\ell^b_i (\pi) +5\ell^b_i (K) )\bigg \}$$

$$\Xi^{*-} \to \Xi^{-} \gamma$$
$$g^{L}_{i} = 2 ~\frac{{\cal C}}{\sqrt{3}} \bigg\{ -(D-F) (\ell^a_i (\pi) - \ell^a_i (K)) + \frac{1}{3} {\cal H} (\ell^b_i (\pi) - \ell^b_i (K) )\bigg \}$$

\newpage

where

\begin{eqnarray}
\ell^a_1 (P) &=& r_p \int^1_0 (1-x) I_{fin} (-x , c) \nonumber \\
\ell^b_1 (P) &=& r_p \int^1_0 (1-\frac{1}{3}x) I_{fin} (x, c) \nonumber \\
\ell^a_2 (P) &=& r_p \int^1_0 (1-2x) I_{fin} (-x, c) ~dx \\
\ell^b_2 (P) &=& r_p \int^1_0 \frac{1}{3} (1-2x) I_{fin} (x, c) ~dx . \nonumber
\end{eqnarray}

\bi

We include the $r_p$ factor to deal with the $f^2_k - f^2_\pi$ difference,
$r_\pi =1$, $r_{_K} =\frac{1}{(1.22)^2}$. This completes the one loop 
calculations, now we turn to the counterterms.

\bi
\bi

\noi {\bf 4b) Counterterms}

\bi
\bi

It is clear that 1-loop calculations reproduces the most general amplitude 
Eq. (5). The counterterms for this calculation can be read by comparing (5) 
with (24,26). To ${\cal O} \bigg( \frac{w^2}{\Lambda^2_\chi} \bigg)$ three 
counterterms contributes to $\frac{3}{2} \to \frac{1}{2} \gamma$

\bi

\begin{eqnarray}
{\cal L}^1_{ct} &=& e ~\frac{\Theta_1}{\Lambda_\chi} \bar B S^{^{\theta}} Q T^\mu F_{_{\theta \mu}} \nonumber \\
{\cal L}^2_{ct} &=& -ie ~\frac{\Theta_2}{\Lambda^2_\chi} \bar B v^\alpha S^\theta Q T^\mu \partial_{\alpha} F_{_{\theta \mu}}  \\
{\cal L}^3_{ct} &=& ie ~\frac{\Theta_3}{\Lambda^2_\chi} \bar B v^\theta S^\alpha Q T^\mu \partial_\alpha F_{_{\theta \mu}}  \nonumber
\end{eqnarray}

\bi

\noi where $Q$ denotes the SU(3) charge matrix $Q= \frac{1}{3} ~{\rm diag} 
~(2, -1, -1)$.  The $SU(3)$ structures of the counterterms are discussed in the
appendix. Notice that ${\cal L}^1_{ct}$ is ${\cal O} \bigg(\frac{w}{\Lambda_\chi}\bigg)$ while ${\cal L}^2_{ct} ~{\rm y} ~{\cal L}^3_{ct}$ are ${\cal O} \bigg(\frac{w^2}{\Lambda^2_\chi}\bigg)$.

\bi

>From (24,26) we see that the 1-loop contributions are ${\cal O} \bigg(
\frac{w^2}{\Lambda^2_\chi}\bigg)$, which means that $\Theta_1$ is finite while
$\Theta_2$ and $\Theta_3$ are infinite constants which renormalize the 1-loop 
contributions.
 
\bi

Including both, one-loop and counterterm contributions, we finally obtain:

\begin{equation}
i {\cal M} = e \bar B \bigg\{ \frac{g_1}{\Lambda^2_\chi} I^1_{\mu\nu\theta} + 
\frac{g_2}{\Lambda^2_\chi}I^2_{\mu\nu\theta}\bigg\}S^\theta T^\mu \varepsilon^\nu
\end{equation}

\noi where

\begin{eqnarray}
g_1 &=& \bigg( \frac{\Lambda_\chi}{w} \Theta_1 + \Theta^R_2 \bigg) Q_{TB} +
g^L_1 \nonumber \\
g_2 &=& \Theta^R_3 ~Q_{TB} + g^L_2 
\end{eqnarray}

\noi $Q_{TB}$ stand for the Clebsch-Gordon coefficientes obtained from Eq. 
(A8), and $\Theta^R_i ~i = 2,3$ are the renormalized constants. By choosing
the normalization scale $\Lambda$ = $\Lambda_\chi$ in Eq. (2), and comparing 
(5), (25) and (32) we obtain:

\begin{equation}
g^D_1 = \frac{2 w}{\Lambda_\chi} g_1 \qquad\qquad g^D_2 = 8 g_2 .
\end{equation}

\noi To leading order in the $\frac{1}{m}$ expansion, from Eq. (2) we obtain 
the multipole amplitudes and decay widths:

\begin{eqnarray}
M1 &=& \frac{e}{3} ~\frac{w^{3/2}}{\Lambda^2_\chi} \left[ 2 g_1 - g_2 \right] \nonumber \\
E2 &=& ~\frac{e}{3} ~\frac{w^{3/2}}{\Lambda^2_\chi} ~g_2 \\
\frac{E2}{M1} &=& ~\frac{g_2}{2 g_1 - g_2} \nonumber \\
\Gamma &=& \frac{2\alpha}{9} \bigg( \frac{w}{\Lambda_\chi} \bigg)^4 w \{ |2g_1 -g_2|^2 + 3 |g_2|^2 \} \nonumber
\end{eqnarray}

\noi  where $\alpha$ denotes the fine structure constant and $w =\Delta m$.

\bi

\noi It is worth-remarking that the leading contributions to $g_1$ are 
${\cal O}\bigg(\frac{w}{\Lambda_\chi}\bigg)$ while, the contributions to $g_2$
are ${\cal O}\bigg(\frac{w^2}{\Lambda^2_\chi}\bigg)$. Thus the $E2$ amplitude 
is subleading in the chiral expansion.

\bi
\bi

\noi {\bf 5. Numerical Results}

\bi

Using HBChPT we have calculated the contributions to the form factors to 
leading order in $\frac{1}{m}$ and to ${\cal O}\bigg(\frac{w^2}{\Lambda^2_\chi}\bigg)$ in the chiral expansion. The $g_1$, $g_2$ form factors 
depend on the low energy constants D,F, ${\cal H}$ and ${\cal C}$ and on the 
renormalization constants $\Theta_1, \Theta^R_2 ~{\rm y} ~\Theta^R_3$. In 
order to make some predictions for the radiative decays of the decuplet it is 
necessary to fix these parameters. The low energy constants $D,F,{\cal H}$ 
~and ${\cal C}$ have been estimated from hyperonic semileptonic
and strong decays [2,3] of the decuplet. The values obtained are 
$|{\cal C}|= 1.6$ , $F=0.40 \pm 0.05$, $D=0.61 \pm 0.04$ and ${\cal H}= -1.9 
\pm 0.7$ wich are in good agreement with the relations obtained from an SU(6) 
symmetry for baryons, namely, $F= \frac{3}{2} D, ~~{\cal C}=-2D, 
~~{\cal H}=-3D$. For numerical calculations we consider isospin as a good 
symmetry. SU(3) breaking is introduced by taking the physical values for  
$f_\pi$ and $f_K$ and also through the mass difference between the isospin 
multiplets. According to our assumptions we take $w = \Delta m = M-m$ 
where $M$ and $m$ is stand for the average mass of the isospin multiplets of 
the decuplet and octet respectively. 

\bi

The ~$SU(3)$ ~forbidden ~decay ~$\Sigma^{*-} ~\to ~\Sigma ~\gamma$ ~and 
~$\Xi^{*-} ~\to ~\Xi^- ~\gamma$, ~for ~wich ~$Q_{_{TB}}=0$, do not receive 
contributions from the counterterms, and therefore are enterely fixed by loop 
calculations. Phenomenologically these decay are very important since they are
induced by SU(3) breaking due to mass terms. For the SU(3) forbiden decays we 
obtain the ratio of multipole amplitudes and braching ratios shown in table 2. 
The numerical calculations are carried using the central values for 
$F,D,{\cal C}$ and ${\cal H}$ obtained in [2,3].

\newpage

\begin{centerline}
{\bf Table 2}
\end{centerline} 

\bi
\bi

\noi Process \hspace{4.8cm} $E2/M1$ \hspace{4cm} $B ~R ~(Exp)$

\bi

\noi $\Sigma^{*-} \to \Sigma^- \gamma$ \hspace{3.5cm} - 0.05 - 0.04 i \hspace{3cm} 0.024 \% ~(?)

\bi

\noi $\Xi^{*-} \to \Xi^- \gamma$ \hspace{3.2cm} -0.05 - 0.06 i \hspace{3cm} ~0.13 \% $(<4\%)$

\bi

The values we obtain for the branching ratios of the $SU(3)$ forbidden decays 
differs by one order of magnitude from those gives in [4] and are closer to 
the experimental bounds [16]. We expect these branching ratios to be correct 
within a factor of 2 due to $\frac{\Delta m}{m}$ corrections.

\bi

For the SU(3) allowed radiative decays of the decuplet, besides the low energy 
constants $D,F,{\cal C},{\cal H}$ it is necessary to fix 
$\Theta\equiv\frac{\Lambda_\chi}{k\cdot v}\Theta_1 +\Theta^R_2$ and 
$\Theta^R_3$, since $\Theta_1$ and $\Theta^R_2$ enter always in the $\Theta$ 
combination. However  existing experimental information 
is not enough to fix these constants. In ref. [1] the value of $\Theta_1$ has 
been fixed from $\Delta\to N\gamma$ by disregarding the contributions from 
$\Theta^R_2 , \Theta^R_3$ under the argument that loop contributions are 
enhanced by the chiral logaritm. We do not agree with this procedure as 
according to the phylosophy of Chiral Perturbation Theory all terms 
contributing to a given order of a process must be included. Futhermore, there
are counter examples to the argument of chiral logs enhancement [15]. In table
3 we quote the predictions we get for the loop contributions to the form 
factors. The numerical calculations are similar to the ones performed for the
forbidden decays (table 2). 

\bi
\bi

\begin{centerline}
{\bf Table 3}
\end{centerline}

\bi

Process \hspace{3.8cm} $g^L_1$ \hspace{4.5cm} $g^L_2$ 

\bi 

\noi $\Delta \to N \gamma$ \hspace{3.3cm} 17.55 - 0.67 $i$ \hspace{3.2cm} 1.29 + 1.98 $i$ 

\bi

\noi $\Sigma^{*0} \to \Lambda \gamma$ \hspace{3cm} -18.05 + 0.39 $i$ \hspace{3cm} -1.07 - 1.31 $i$

\bi

\noi $\Sigma^{*0}\to \Sigma^0 \gamma$ \hspace{3cm} 20.13 \hspace{4.6cm}  0.15 

\bi

\noi $\Sigma^{*+} \to \Sigma^+ \gamma$ \hspace{2.7cm} -37.28 + 0.02 ~$i$ \hspace{2.7cm} -0.63 - 0.22 $i$

\bi

\noi $\Sigma^{*-} \to \Sigma^- \gamma$ \hspace{2.9cm} 2.97 + 0.02 $i$ \hspace{3cm}-0.33 - 0.22 $i$

\bi

\noi $\Xi^{*-} \to \Xi^- \gamma$ \hspace{2.9cm} 2.58 + 0.05 $i$ \hspace{3cm}-0.29 - 0.30 $i$

\bi

\noi $\Xi^{*0} \to \Xi^0 \gamma$ \hspace{2.8cm} -33.13 + 0.05 $i$ \hspace{3cm}-0.64 - 0.30 $i$

\bi
\bi  

\begin{centerline}
{\bf Conclusions}
\end{centerline}

\bi

- We ~analyzed ~the ~interacting ~spin 3/2 field theory in the light of 
R-invariance, which is necessary in order to ensure the non-physical nature of the spin
$\frac{1}{2}$ content of the Rarita-Schwinger spinor. We have seen that 
R-invariance can be implemented at the price of introducing the so called  
``off-shell" parameters.

- We have argued that to order zero in the heavy baryon expansion, terms 
involving the arbitrary ``off-shell" parameters do not contribute to the 
S-matrix elements. Even more, to the same order in the heavy baryon expansion,
the subsidiary conditions reduces to those of the free theory, ensuring that  
the interacting theory involves only the $S= \frac{3}{2}$ degrees of freedom.

\bi

- The interacting spin 3/2 theory is free of the ``off-shell" ambiguities only
to order zero in the heavy baryon expansion. This implies that the heavy 
baryon and the chiral expansion must be separately considered even if the 
expansion parameters, the baryon mass $m \approx 1 GeV$ and $\Lambda_\chi 
\approx 1 GeV$ respectively, are of the same order. 

\bi

- We applied the formalism previously described to the calculation of 
radiative decays of the spin 3/2 decuplet. We performed an explicitly gauge 
invariant calculation wich permits to unambiguously identify the form factors 
as well as the counterterms required by the chiral expansion. The calculation 
is valid to leading order in $\frac{1}{m}$ and to ${\cal O} \bigg(\frac{w^2}{\Lambda^2_\chi}\bigg)$ in the chiral expansion. We present predictions for the 
$SU(3)$ forbidden decays $\Sigma^{*-}\to\Sigma^- \gamma$ and $\Xi^{*-}\to\Xi^-
\gamma$. Our results are one order of magnitude larger than previous estimates
[4]. We also report analytical and numerical results for the 1-loop 
contributions to the $SU(3)$ allowed radiative decays of the decuplet.
Unfortunately we are not able to present predictions for the latter decays 
since it is necesary to fix two low energy constants wich is not 
possible with the available experimental information.

\newpage

  \begin{appendix}

\section*{Appendix}
\newcommand{\sect}[1]{\setcounter{equation}{0}
 \renewcommand{\theequation}{#1.\arabic{equation}}
}
\sect{A}

\bi

The lowest order Heavy Baryon Lagrangian for Spin 1/2 Fields is [3]

\bq\label{(A1)}
{\cal L}^{(1)}_8 &=& <\bar B (v\cdot i\partial)B+2iD \bar B S^\mu \{\Delta_\mu ,B\}+ 2iF \bar B S^\mu [\Delta_\mu , B] +  \\ 
&~&\frac{f^2}{4} D_\mu\Sigma D^\mu\Sigma^+ + a{\cal M}(\Sigma +\Sigma^+)>\nonumber
\eq

where $< >$ denotes trace over $SU(3)$ structure, ${\cal M}$ denotes the quark
mass matrix and B is a matrix containing the spin 1/2 baryon fields. Altough 
not explicitly written throughout this section baryon fields denotes definite 
velocity fields. Pseudoscalar fields enter in the formalism through the 
matrices $\Phi , ~\xi \equiv \exp \bigg(\frac{i}{f} \Phi \bigg)$ and
$\Sigma \equiv  \xi^2$ where $\Phi$ contains the pseudogoldstone fields.
Explicitly 

\bi

$$\Phi= \frac{1}{2} \lambda_\alpha \phi^\alpha = \frac{1}{\sqrt{2}} \pmatrix{\frac{\pi^0}{\sqrt{2}} + \frac{\eta}{\sqrt{6}}&\pi^+&K^+\cr
\pi^-&-\frac{\pi_0}{\sqrt{2}} + \frac{\eta}{\sqrt{6}}&K^0\cr
K^-&\bar K^0&-\frac{2 \eta}{\sqrt{6}}\cr},$$

\be\label{(A2)}
B= \frac{1}{\sqrt{2}} \lambda_\alpha B^\alpha = \pmatrix{\frac{\Sigma^0}{\sqrt{2}} + \frac{\Lambda}{\sqrt{6}}&\Sigma^+&p\cr
\Sigma^-&-\frac{\Sigma^0}{\sqrt{2}} + \frac{\Lambda}{\sqrt{6}}&n\cr
\Xi^-&\Xi^0&-\frac{2 \Lambda}{\sqrt{6}}\cr}, 
\en

$$\bar B = \frac{1}{\sqrt{2}} \lambda_\alpha \bar B^\alpha = \pmatrix{\frac{\bar\Sigma^0}{\sqrt{2}} + \frac{\bar \Lambda}{\sqrt{6}},&\bar\Sigma^-,&\bar\Xi^-\cr
\bar\Sigma^+,&-\frac{\bar\Sigma^0}{\sqrt{2}} + \frac{\bar\Lambda}{\sqrt{6}}&\bar\Xi^0\cr
\bar p&\bar n&-\frac{2 \bar\Lambda}{\sqrt{6}}\cr}.$$   

\bi

With these conventions we have $f_\pi \approx 93 ~MeV$. The chiral covariant 
derivatives contains the chiral conexion 

\bi

\be\label{(A3)}
{\cal D}_\mu B = \partial_\mu B + [\Gamma_\mu, B] . 
\en

\bi

\noi This conexion and the chiral covariant axial vector field $\Delta_\mu$, 
are given as 

\bi

\bq\label{(A4)}
\Gamma_\mu &=& \frac{i}{2} \{ \xi \partial_\mu \xi^+ + \xi^+ \partial_\mu \xi \} = \frac{1}{2 f^2} [\Phi, \partial^\mu \Phi ] + \cdots \\
\Delta_\mu &=& \frac{1}{2} \{ \xi \partial_\mu \xi^+ - \xi^+ \partial_\mu \xi\} = \frac{1}{f} \partial^\mu \Phi - \frac{1}{6 f^3} [\Phi, [\Phi, \partial_\mu 
\Phi]] + \cdots \nonumber
\eq

\bi

Interactions with external sources are introduced through covariant derivatives
wich for electromagnetic interactions reduces to the substitutions 

\begin{eqnarray}
\Gamma_\mu &\to& \Gamma_\mu + \frac{i e}{2} A_\mu (\xi Q \xi^+ + \xi^+ Q \xi) \nonumber \\
\Delta_\mu &\to& \Delta_\mu - \frac{e}{2} A_\mu ( \xi Q \xi^+ - \xi^+ Q \xi) \nonumber \\
{\it D}_\mu \Sigma &=& \partial_\mu \Sigma + i e A_\mu [Q, \Sigma] . \nonumber
\end{eqnarray}

\bi

If we consider one-meson exchange only this reduces to

\bi

$$\Gamma_\mu \to i e Q A_\mu \qquad\qquad\quad$$
\be\label{(A5)}
\Delta_\mu \to \frac{1}{f}\partial_\mu\Phi +\frac{ie}{f} A_\mu [Q,\Phi]
\en 
$${\cal D}_\mu B = \partial_\mu B + i e A_\mu [Q,B] . \qquad\quad $$

\bi

\noi According to the main text, if we work to leading order in the 
$\frac{1}{m}$ expansion we can forget those terms in Eqs. (12,18) containing
the ``off-shell" parameters. The resulting lagrangian for the spin 3/2 sector
is 

\bq\label{(A6)}
{\cal L}_{10}&=& \bar T^\mu (i \partial_\alpha \Gamma^\alpha_{\mu\nu} - M B_{\mu\nu} )_{HB} T^\nu + i{\cal C}(\bar T^\mu \Delta_\mu B +\bar B \Delta_\mu 
T^\mu) \\
&+& 2 i {\cal H} \bar T^\mu S_\nu \Delta^\nu T_\mu \nonumber
\eq

\noi where
 
$$({\cal D}^\nu T^\mu)_{abc} \equiv \partial^\nu T^\mu_{abc} + (\Gamma^\nu)_a\,^d T^\mu_{dbc} + (\Gamma^\nu)_b\,^d T^\mu_{adc} + (\Gamma^\nu)_c\,^d T^\mu_{abd} .$$

\bi

The $SU(3)$ structures are 

\bi

\be\label{(A7)}
\bar T \Delta B + h.c \equiv \varepsilon^{kmn} \bar T_{ijk} \Delta^i\,_m B^j\,_n + h.c 
\en
$$\bar T \Delta T \equiv \bar T_{jk\ell} (\Delta )^j\,_m T^{mk\ell} .$$

\bi

We use the subindex $HB$ in Eq. (A6) to denote the Heavy-Baryon limit of the
first term. According to the main text as far as we consider one meson 
exchange and leading order terms in the $\frac{1}{m}$ expansion this term 
reduces to a free term wich produces the propagator given in Eq. (19) plus a 
minimally coupled electromagnetic interaction. 

\bi

$${\cal L}_{TAT} =i T^\mu v^\alpha (ie Q A_\alpha) T_\mu =-e v\cdot A T^{abc}_{\mu} (QT^\mu)_{abc}$$

\noi where

\bi

$$(QT)_{abc} \equiv Q_a\,^d T_{dbc} + Q_b\,^d T_{adc} + Q_c\,^d T_{abd} .$$

\bi

Physical spin 3/2 fields and SU(3) fields are related by

\bi

$$T_{111}= \Delta^{++} ~, ~~T_{112} = \frac{1}{\sqrt{3}} \Delta^+ ~, ~~T_{122} = \frac{1}{\sqrt{3}} \Delta^0 ~, ~~T_{222} = \Delta^-$$
$$T_{113} = \frac{1}{\sqrt{3}} \Sigma^{++} ~, ~~T_{123} = \frac{1}{\sqrt{6}} \Sigma^{* 0} ~, ~~T_{223} = \frac{1}{\sqrt{3}} \Sigma^{*-}$$
$$T_{123} = \frac{1}{\sqrt{3}} \Xi^{*0} ~, ~~T_{233} = \frac{1}{\sqrt{3}} \Xi^{*-}$$
$$T_{333} = \Omega^- .$$

\bi

In addition to the leading order terms we have the counterterms given in Eq. 
(31)  wich have the $SU(3)$ structure

\be\label{(A8)}
\bar B QT \equiv \varepsilon^{kmn} \bar B_m\,^i Q^j\,_n T_{ijk} .
\en

\bi

Feynman rules we require to perform the calculation of the non-nule diagrams 
shown in fig. 1 are 

\bi

$$\qquad\qquad =\frac{\sqrt{2}i}{f} (C.G.)_{B B\pi} S\cdot\ell \hspace{4cm} \frac{\sqrt{2}i}{f} (C.G)_{TT\pi} (S\cdot\ell) g_{\alpha\beta}$$ 

\bi

$$\qquad\qquad \frac{i}{\sqrt{2}f} (C.G)_{TB\pi} \ell_\mu \hspace{4.5cm} -iq(\ell_1 +\ell_2)_\mu$$

\bi

$$ie \frac{\Theta_1}{\Lambda_\chi} Q_{_{TB}} \frac{1}{k\cdot v} I^1_{\mu\nu\theta} S^\theta$$
$$ie \frac{\Theta_2}{\Lambda^2_\chi} Q_{_{TB}} I^1_{\mu\nu\theta} S^\theta$$
$$ie \frac{\Theta_3}{\Lambda^2_\chi} Q_{_{TB}} I^2_{\mu\nu\theta} S^\theta$$

\bi

where double (single) lines denotes the spin 3/2 (spin 1/2) fields, dashed  
(wavy) lines denotes the pseudoscalar (photon) field and $q, (C.G)$ denotes 
the charge of the pseudoscalar entering in the loop and the product of 
Clebsch-Gordon coefficients respectively.

\bi

\noi With these conventions we obtain the same values for the products of 
$(C.G)$ coefficients and for $Q_{_{TB}}$ as those listed in ref [4].

\end{appendix}

\newpage

 \begin{center} {\large \bf REFERENCES}
\end{center}
\small
\begin{itemize}
\item[1.-] For a review see ``Dynamics of the standard model" J.F. Donoghue, 
	   E. Golowich and B.R. Holstein, Cambridge Univ. Press (1992).
\item[2.-] E. Jenkins \& A.V. Manohar Phys. Lett. B255 (1991), 558.
\item[3.-] E. Jenkins \& A.V. Manohar Phys. Lett. B259 (1991), 353. For a 
	   review see ``BCHPT" E. Jenkins and A. Manohar contributions to 
	   Workshop on Effective Field Theories of the Standard Model. Ed. U. 
	   Meissner (World Scientific, Singapore, 1992.)
\item[4.-] M.N. Butler {\it et.al.} Nucl. Phys. B399 (1993), 69.  
\item[5.-] N. Isgur {\it et.al.} Phys. Rev. D25, (1982), 2394 and references
	  there in.
\item[6.-] W. Rarita \& J. Schwinger. Phys. Rev. 60, 61 (1941)
\item[7.-] C. Fronsdal Nuovo Cimento (suppl.) IX, 416 (1958). J. Urias, Ph D. 
	   Thesis, Cat\'olique University of Louvain, Belgium (1976), 
	   unpublished. 
\item[8.-] L.M. Nath, {\it et. al.} Phys. Rev. D3, 2153 (1971);Z. Phys. C5 
	   (1980), 9.
\item[9.-] R.M. Davidson {\it et. al.} Phys. Rev. D43, (1991), 71. 

	   F. Jones \& M.D. Scadron, Ann. Phys. 81, 1, 1973.
\item[10.-] S. Kamefuchi, L. O'Raifeartaigh and A. Salam Nucl. Phys. 28, 
	   (1961), 529.
\item[11.-] M.G. Olsson and E.T. Ossypowski, Nucl. Phys. B87 (1975), 339; 
	    Phys. Rev. D17 (1978), 174.
\item[12.-] M. Napsuciale and J.L. Lucio M. hep-ph/9605266. To be published 
	    in Phys. Lett. B.; M. Napsuciale Ph. D. Thesis CINVESTAV, M\'exico
	    (1995).
\item[13.-] R.D. Peccei, Phys. Rev. 181 (1969) 1902; Phys. Rev. 176 (1968)
	    1812.
\item[14.-] 't Hooft and M. Veltman Nucl. Phys. B153 (1979), 365. 
\item[15.-] See page 2387 of J.F. Donoghue and B.R. Holstein Phys. Rev. D40 
	   (1989) 2378.
\item[16.-] Review of Particles Properties Phys. Rev. D50 (1994).
\end{itemize} 
\end{document}